# Electron transport in GaN(ZB) and AlN(WZ)

Clóves G. Rodrigues

In the recent years have been increasing the interest on the electronic properties of the III-nitride semiconductors due to their promising technological applications such as blue-yellow light emitting diode [1]. The optoelectronic device prototypes based in the III-nitride heterostructures operating under external electric fields are quite common [1]. In that case, the carriers are in far-from-equilibrium conditions, and consequently it is necessary to resort to appropriate nonequilibrium statistical mechanical and thermodynamic formalisms for, e.g., the study of their optical and transport properties.

The transport phenomena in semiconductors have been studied using balance equation approach [2], Green's-function approach [3], Boltzmann transport equations [4] and Monte Carlo simulations [5]. However, methods based on Boltzmann-like balance transport theories have limitations when nonlinear effects become important. The computational modeling, such as Monte Carlo approach, gives a reasonable good agreement with experimental data, but the method used here has the advantage that it provides analytical equations and a better interpretation of the results. The method used in this work is a nonlinear quantum kinetic theory based on a nonequilibrium ensemble formalism, so-called *Non Equilibrium Statistical Operator Method* (NESOM) [6]. The NESOM is a powerful formalism that seems to offer an elegant and concise way for an analytical treatment in the theory of irreversible processes, adequate to deal with a large class of experimental situations.

Seeking a better understanding of the electron transport in n-doped zinc blende GaN and wurtzite AlN, it was performed here a theoretical study resorting to NESOM, which provides a set of coupled nonlinear differential evolution equations that provides a description of the dissipative phenomena developed in the system. The ultrafast time evolution (and the steady-state) of the electron drift velocity is obtained assuming different values for the electron effective mass.

Let us consider now the nonequilibrium thermodynamic state of the single plasma: the presence of the electric field $\mathbf{F}$ changes the energy of the electrons (they acquire energy in excess of equilibrium), and these carriers keep transferring this excess to the lattice, and an electrical current follows it. Thus, we need to choose as basic variables: the energy of the electrons $E_e(t)$, the linear momentum of the electrons $\mathbf{P}(t)$, the number of the electrons $N(t)$, the energies of the longitudinal optical phonons $E_{lo}(t)$ and the energies of the acoustical phonons $E_{ac}(t)$. The transversal optical phonons are ignored, once they weakly interact with the electrons in the conduction band [7].

The evolution equations these basic variables are derived using a nonlinear quantum transport theory (more details given in the Ref. [8]). They are

$$\frac{d}{dt}E_e(t) = \frac{e\mathbf{F}}{m^*} \cdot \mathbf{P}(t) - J^{(2)}_{E_e}(t), \qquad (1)$$

$$\frac{d}{dt}\mathbf{P}(t) = Ne\mathbf{F} - \mathbf{J}^{(2)}_{\mathbf{P},\text{ph}}(t) - \mathbf{J}^{(2)}_{\mathbf{P},\text{imp}}(t), \qquad (2)$$

C. G. Rodrigues (✉)
Núcleo de Pesquisa em Física, Departamento de Física,
Universidade Católica de Goiás, CP 86, 74605-010 Goiânia,
Goiás, Brazil
cloves@pucgoias.edu.br

$$\frac{d}{dt}E_{\text{lo}}(t) = J^{(2)}_{\text{lo}}(t) - J^{(2)}_{\text{lo,an}}(t), \tag{3}$$

$$\frac{d}{dt}E_{\text{ac}}(t) = J^{(2)}_{\text{ac}}(t) + J^{(2)}_{\text{lo,an}}(t) - J^{(2)}_{\text{ac,dif}}(t), \tag{4}$$

$$\frac{d}{dt}N(t) = 0, \tag{5}$$

where $m^*$ is the electron effective mass, $e$ is the electron charge, and $\mathbf{F}$ is the electric field. It should be noticed that all the collision operators $J^{(2)}(t)$ are dependent on the intensive nonequilibrium thermodynamic variables: the so-called quasitemperatures (sometimes referred by some authors as nonequilibrium temperature) $T_e^*(t)$, $T_{\text{lo}}^*(t)$, $T_{\text{ac}}^*(t)$, of electrons and phonons, the quasi-chemical potential, $\mu^*(t)$, and the electron drift velocity, $\mathbf{v}(t)$. The external reservoir is assumed to remain at a constant temperature $T_R$. To close the system of equations we need to express the energy, linear momentum of the electrons, the energies of the $lo$ and $ac$ phonons, in term of the former, that is:

$$E_e(t) = \sum_{\mathbf{k}}(\hbar k^2/2m^*)f_{\mathbf{k}}(t), \tag{6}$$

$$\mathbf{P}(t) = \sum_{\mathbf{k}} \hbar \mathbf{k} f_{\mathbf{k}}(t), \tag{7}$$

$$E_{\text{lo}}(t) = \sum_{\mathbf{q}} \hbar \omega_{\mathbf{q},\text{lo}} v_{\mathbf{q},\text{lo}}(t), \tag{8}$$

$$E_{\text{ac}}(t) = \sum_{\mathbf{q}} \hbar \omega_{\mathbf{q},\text{ac}} v_{\mathbf{q},\text{ac}}(t), \tag{9}$$

$$N(t) = \sum_{\mathbf{k}} f_{\mathbf{k}}(t), \tag{10}$$

where $\omega_{\mathbf{q},\text{lo}}$ is the frequency of the $lo$ phonons and $\omega_{\mathbf{q},\text{ac}}$ is the frequency of the $ac$ phonons. In this equations $f_{\mathbf{k}}(t)$ is the time-dependent (on the evolution of the nonequilibrium state of the system) distribution's function for electrons (a drifted Maxwell–Boltzmann-like distribution)

$$f_{\mathbf{k}}(t) = n\left(\frac{2^{1/3}\pi\hbar^2}{m^* k_B T_e^*(t)}\right)^{3/2} \exp\left\{-\frac{\hbar^2(\mathbf{k} - m^*\mathbf{v}(t)/\hbar)^2}{2m^* k_B T_e^*(t)}\right\}, \tag{11}$$

where $k_B$ is the Boltzmann constant. Moreover, $v_{\mathbf{q},\eta}(t)$ is the populations of the phonons:

$$v_{\mathbf{q},\eta}(t) = [\exp\{\hbar\omega_{\mathbf{q},\eta}/k_B T_\eta^*(t)\} - 1]^{-1}, \tag{12}$$

where $\eta = lo$, $ac$ (for longitudinal optical and acoustical phonons respectively). We notice that we have used the Debye approximation $(\omega_{\mathbf{q},\text{ac}} \simeq v_s q)$ for the acoustic branch and Einstein approximation $(\omega_{\mathbf{q},\text{lo}} \simeq \omega_{\text{lo}})$ for the optical branch.

Let us analyze the right side of the Eqs. 1–5 term by term. In Eq. 1 the first term on the right accounts for the rate of energy transferred from the electric field to the carriers. The second term accounts for the transfer of the excess energy of the carriers—received from the first term—to the $lo$ and $ac$ phonons

$$J^{(2)}_{E_e} = \frac{2\pi}{\hbar} \sum_{\mathbf{k},\mathbf{q},\ell,\eta} |M_\eta^\ell(\mathbf{q})|^2 (\epsilon_{\mathbf{k}+\mathbf{q}} - \epsilon_{\mathbf{k}})[v_{\mathbf{q},\eta}(t)f_{\mathbf{k}}(t) \\ \times (1 - f_{\mathbf{k}+\mathbf{q}}(t)) - f_{\mathbf{k}+\mathbf{q}}(t)(1 + v_{\mathbf{q},\eta}(t)) \\ \times (1 - f_{\mathbf{k}}(t))]\delta(\epsilon_{\mathbf{k}+\mathbf{q}} - \epsilon_{\mathbf{k}} - \hbar\omega_{\mathbf{q},\eta}), \tag{13}$$

where $\epsilon_{\mathbf{k}} = \hbar^2 k^2/2m_e^*$ and $\eta = lo$ and $ac$. The quantity $M_\eta^\ell(\mathbf{q})$ is the matrix element of the interaction between carriers and $\eta$-type phonons, with supraindex $\ell$ indicating the kind of interaction (polar, deformation potential, piezoelectric).

In Eq. 2 the first term on the right is the driving force generated by the presence of the electric field. The second term is the rate of momentum transfer due to interaction with the phonons given by

$$\mathbf{J}^{(2)}_{\mathbf{P},ph}(t) = 2\pi \sum_{\mathbf{k},\mathbf{q},\ell,\eta} \mathbf{q}|M_\eta^\ell(\mathbf{q})|^2[v_{\mathbf{q},\eta}(t)f_{\mathbf{k}}(t) \\ (1 - f_{\mathbf{k}+\mathbf{q}}(t)) - f_{\mathbf{k}+\mathbf{q}}(t)(1 + v_{\mathbf{q},\eta}(t))(1 - f_{\mathbf{k}}(t))] \\ \times \delta(\epsilon_{\mathbf{k}+\mathbf{q}} - \epsilon_{\mathbf{k}} - \hbar\omega_{\mathbf{q},\eta}), \tag{14}$$

where, we recall, $\eta = lo$ and $ac$. The third term in Eq. 2 is the scattering due to impurities:

$$\mathbf{J}^{(2)}_{\mathbf{P},\text{imp}} = \mathbf{P}(t)/\tau_{\text{imp}}(t), \tag{15}$$

using for $\tau_{\text{imp}}(t)$ the expression given by Ridley adapted from the results reported by Brooks–Herring [9]:

$$\tau_{\text{imp}}(t) \simeq \frac{128\sqrt{2\pi m_e^*}(k_B T_e^*(t))^{3/2}}{\mathcal{N}_I(\mathcal{Z}e^2/\epsilon_0)^2 G(t)}, \tag{16}$$

where $\mathcal{N}_I$ is the density of impurities, $\mathcal{Z}$ the units of charge of the impurity, and

$$G(t) = \ln[1 + b(t)] - \frac{b(t)}{1 + b(t)}, \tag{17}$$

where

$$b(t) = \frac{24\epsilon_0 m_e^*(k_B T_e^*(t))^2}{\mathcal{N}_I e^2 \hbar^2}. \quad (18)$$

In Eqs. 3 and 4 the first term on the right describes the rate of change of the energy of the phonons due to interaction with the electrons. More precisely they account for the gain of the energy transferred to them from the hot carriers and then contribution $J_{lo}^{(2)}(t)$ and $J_{ac}^{(2)}(t)$ are the same as those calculated in Eq. 13 for $\eta = lo$ and $\eta = ac$, with a change of sign. The second term in Eq. 3 accounts for the rate of transfer of energy from the optical phonons to the acoustic ones, given by

$$J_{lo,an}^{(2)}(t) = \sum_{\mathbf{q}} \hbar\omega_{\mathbf{q},lo} \frac{v_{\mathbf{q},lo}(t) - v_{\mathbf{q},lo}^{ac}}{\tau_{lo,an}}, \quad (19)$$

where

$$v_{\mathbf{q},lo}^{ac}(t) = \frac{1}{\exp\{\hbar\omega_{\mathbf{q},lo}/k_B T_{ac}^*(t)\} - 1}, \quad (20)$$

$\tau_{lo,an}$ being a relaxation time which is obtained from the band width in Raman scattering experiments, as in Ref. [10]. The contribution $J_{lo,an}^{(2)}(t)$ is the same but with different sign in Eqs. 3 and 4, that is, the energy is transferred from the optical phonons to the acoustical ones.

The last term in Eq. 4 is the diffusion of heat from the acoustical phonons to the reservoir at a temperature $T_R$

$$J_{ac,dif}^{(2)}(t) = \sum_{\mathbf{q}} \hbar\omega_{\mathbf{q},ac} \frac{v_{\mathbf{q},ac}(t) - v_{\mathbf{q},ac}^{eq}}{\tau_{ac,dif}}, \quad (21)$$

where $\tau_{ac,dif}$ is a characteristic time for heat diffusion, which depends on the particularities of the contact of sample and reservoir [11]. Finally, the Eq. 5 accounts for the fact that the concentration $n$ of electrons is fixed by doping.

Here, we solve the evolution Eqs. 1–4 for the n-doped GaN(ZB) and AlN(WZ) with $n = 10^{17}$ cm$^{-3}$ at room temperature, that is, $T_R = 300$ K. Besides, the electric field is applied on the system initially in equilibrium, and then $T_e^*(0) = T_{lo}^*(0) = T_{ac}^*(0) = 300$ K and $\mathbf{v}(0) = 0$. In these calculations were considered the interaction between phonons and carriers (Fröhlich, acoustic deformation, and piezoelectric) and also the scattering by ionized impurities. The parameters used in calculations for GaN(ZB) and AlN(WZ) are showed in Table 1.

**Table 1** Parameters of AlN(WZ) and GaN(ZB)

| Parameter | AlN(WZ) | GaN(ZB) |
|---|---|---|
| Lattice parameter $a$, (Å) | 3.11 [12] | 4.5 [14] |
| Lattice parameter $c$, (Å) | 4.98 [12] | – |
| $lo$ phonon energy $\hbar\omega_{lo}$, (meV) | 99.2 [12] | 92 [15] |
| Static dielectric constant $\epsilon_0$ | 8.5 [12] | 9.5 [16] |
| Optical dielectric constant $\epsilon_\infty$ | 4.77 [12] | 5.35 [16] |
| Mass density $\rho$, (g/cm$^3$) | 3.23 [12] | 6.09 [17] |
| Long. elastic const. $C_l$, ($\times 10^{12}$dyn/cm$^2$) | 2.65 [12] | 2.66 [18] |
| Trans. elastic const. $C_t$, ($\times 10^{11}$dyn/cm$^2$) | 4.42 [12] | 4.41 [18] |
| Acoustic deform. potential $E_1$, (eV) | 9.5 [12] | 10.1 [19] |
| Piezoelectric constant $h_{pz}$, (C/m$^2$) | 0.92 [13] | 0.56 [20] |

The Fig. 1 shows the time evolution towards the steady state of the electron drift velocity in n-doped AlN(WZ) (Fig. 1a) and GaN(ZB) (Fig. 1b), for four different values of the electron effective mass (Refs. [14, 21–26]) in the presence of an electric field of intensity 80 kV/cm. It is verified that the steady state is attained after a transient time of the order of about 0.25 picoseconds. At examining the Fig. 1, we can notice that the less the electron effective mass, the more the velocity overshoot. The overshoot effect follows if during the time evolution of the macroscopic state of the system, under the action of the electric field, the electron relaxation rate of momentum is larger than the electron relaxation rate of energy. It is verified that the electron drift velocity depends critically on the value of the electron effective mass. For this electric field of intensity 80 kV/cm the electron drift velocity (in the steady state) in GaN(ZB) for $m^* = 0.13 m_0$ is $3.3 \times 10^7$ cm/s and for $m^* = 0.21 m_0$ is $2.0 \times 10^7$ cm/s, that is, a difference of about 39%. For AlN(WZ) the electron drift velocity (in the steady state) for $m^* = 0.25 m_0$ is $1.7 \times 10^7$ cm/s and for $m^* = 0.40 m_0$ is $1.1 \times 10^7$ cm/s, resulting a difference of about 35%.

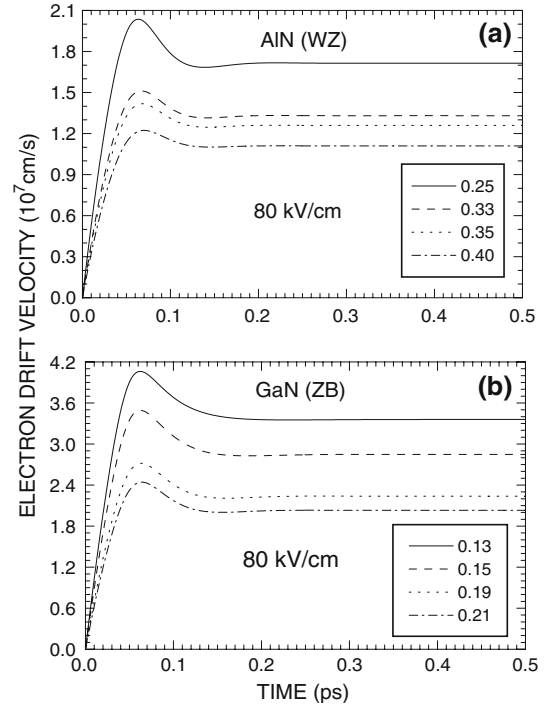

**Fig. 1** The time evolution of the electron drift velocity, for the different values of the electron effective mass: (**a**) AlN(WZ): $m^* = 0.25m_0$, Ref. [21]; $m^* = 0.33m_0$, Ref. [22]; $m^* = 0.35m_0$, Ref. [23]; $m^* = 0.40m_0$, Ref. [24]; (**b**) GaN(ZB): $m^* = 0.13m_0$, Ref. [25]; $m^* = 0.15m_0$, Ref. [26]; $m^* = 0.19m_0$, Ref. [23]; $m^* = 0.21m_0$, Ref. [14], where $m_0$ is the electron rest mass

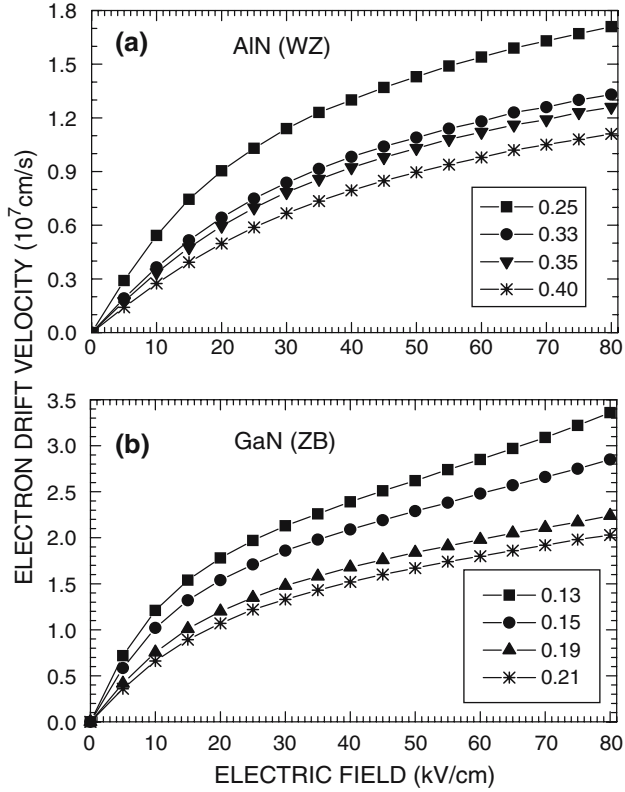

**Fig. 2** The dependence of the electron drift velocity in the steady state with the electric field in n-doped (**a**) AlN(WZ) and (**b**) GaN(ZB), for four different values of the electron effective mass

**Table 2** Mobility of AlN(WZ) and GaN(ZB)

|  | $m^*$ ($m_0$) | Mobility ($cm^2$/V s) |
| --- | --- | --- |
| AlN(WZ) | 0.25 [21] | 599 |
|  | 0.33 [22] | 387 |
|  | 0.35 [23] | 352 |
|  | 0.40 [24] | 285 |
| GaN(ZB) | 0.13 [25] | 1560 |
|  | 0.15 [26] | 1252 |
|  | 0.19 [23] | 870 |
|  | 0.21 [14] | 744 |

The Figs. 2a and b show the dependence of the electron drift velocity (in the steady state) for the two materials with the electric field strength, for four different values of the electron effective mass. It can be seen that at low electric fields, an Ohmic region is present, following a departure from the Ohmic behavior for larger fields.

We can derive the mobility in the steady state, $\mathcal{M}$ of the carriers, as given by $\mathcal{M} = |\mathbf{v}|/|\mathbf{F}|$, with the electron drift velocity $\mathbf{v}$ related to linear momentum per electron by $\mathbf{P} = m^* \mathbf{v}$. The Table 2 shows the mobility (at low field, that is, Ohmic region) for AlN(WZ) and GaN(ZB) for different values of the electron effective mass. It can be noticed that the larger mobility corresponds to a smaller electron effective mass.

The Figs. 1 and 2, and Table 2 show how important is to know the accurate value of the electron effective mass for doing theoretical calculations and simulations of electronic transport in the GaN(ZB) and AlN(WZ), since, the results obtained depend critically on the value assumed for the electron effective mass.